\documentstyle[12pt,openbib]{article}

\begin{document}



\title{Effect of H$^-$ ion formation on Positronium-Hydrogen 
elastic scattering}

\author{P K Biswas \\
Departamento de F\'{\i}sica, Instituto Tecnologico da Aeronautica,
CTA\\
S\~ao Jos\'e dos Campos 12228-900, SP, Brasil\\  }



\maketitle

\begin{abstract}

Effect of charge-transfer recombination for positronium (Ps) scattering
is studied on Ps-H system using coupled-channel formalism considering a
new rearrangement channel Ps+H$\rightarrow e^+$+H$^-$ and exchange. The
correlation and continuum effects introduced by this charge-transfer
channel result to 
substantial convergence to the low energy scattering parameters. Effects
on scattering length, PsH binding energy, and low-energy (0-10 eV) cross
sections are evaluated and H$^-$ formation cross sections are reported
from above the threshold (6.438 eV) to 100 eV.


\end{abstract} 

\vskip 12cm

\newpage {\bf Introduction :} Charge-transfer recombinations have been
found to be of fundamental importance in simulating the continuum effect
in the scattering dynamics of positrons \cite{msa,msb}. Their roles are
yet to be investigated for the scattering of positronium (Ps) atom - an
exotic bound state of electron and positron with singlet and triplet
spins.  This seems to be very important since, despite a long history of
theoretical studies
\cite{mot,hf,dr1,dr2,dr3,bb,wl0,wl1,wl2,hrg,sg1,sg2,ng,hasi,cpl,nimb,pb1,
bam2,ba1,ab1,abs,bam,ba0,bd}, we are yet to produce converged scattering
results in the {\it ab initio} coupled-channel (CC) framework, even for
the simplest Ps-H system \cite{wl1}.

So far, in dealing with Ps scattering, emphasis has been given to the
exchange rearrangement of electrons, which is also of fundamental
importance, as the static-potential for a Ps-target system vanishes due
to internal charge-and-mass symmetry of Ps. However, exchange constitutes
only a part of the continuum and Ps exhibits a strong polarizability of
$36 a_0^3$, which is expected to make it vulnerable to continuum effects
and charge transfer processes.  Continuum effect can be simulated through
pseudostates of Ps and target, but for Ps scattering this is expected
lead to an untractable situation as the number of scattering channels
grows like $N^2$, where $N$-number of intermediate states for each of Ps
and target are considered. Considering the difficulty, so far, the
pseudostate technique has been applied (to Ps-H and Ps-He scattering
problem) employing pseudostates for only Ps. This is not expected to
simulate the appropriate continuum effect as the target is kept idle to
keep the calculational schemes tractable. However, from the experience in
positron-atom scattering \cite{msb}, we understand that allowed charge
transfer processes can also enhance the continuum effect significantly
and lead to the required convergence of the theoretical scheme. The
effect of such charge-transfer channels is yet unexplored for Ps
scattering and is the subject of this work.

Apart from the initial works \cite{mot,hf,dr1,dr2,dr3,bb}, the recent
theoretical development on Ps scattering studies, has contributions from
the Belfast group \cite{wl0,wl1,wl2}, the Calcutta group
\cite{hrg,sg1,sg2} and the S\~ao Paulo group
\cite{cpl,nimb,pb1,bam2,ba1,ab1,abs,bam,ba0,bd}.  While the Belfast and
Calcutta groups use complete {\it ab initio} coupled-channel (CC)
formalism, the S\~ao Paulo group indulges to model exchange correlation
potentials (to enhance the correlation and continuum effects) in the {\it
ab initio} CC formalism. In this investigation, we concentrate on the
simplest but the
richest Ps-H scattering problem, where predictions of the {\it ab initio}
coupled-channel (CC) calculations are yet to converge with those of the
accrrate variational calculations \cite{fs,ho1}.

In the {\it ab initio} CC formalism, the largest continuum effect has
been taken in a 22-state calculation which uses three eigen states and
nineteen pseudostates of Ps keeping the target idle. This calculation
yields PsH binding energy of 0.634 eV and S-wave singlet resonance
energy of 4.55 eV compared to the most accurate variational estimates of
1.067 eV \cite{fs} and 4.003 eV \cite{ho1}, respectively. The lack of
convergence was expected as the virtual effects of the target are
neglected. Interestingly, predictions of this CC calculation come to
close agreement with the predictions of another variational calculation
\cite{dr1} that has been performed without explicit electron-electron
correlation in their trial ansatz. The precise variational calculations
above \cite{fs,ho1} includes explicit $e-e$ correlation in their trial
wave functions and they also contain H$^-$ flavour. These clearly
emphasize the necessity of the introduction of more electron-electron
correlation and continuum effects to the present state of the art CC
calculations. We understand that, for Ps-H system, electron-electron
correlation and continuum effects could be enhanced through the formation
of H$^-$-ion via Ps+H$\rightarrow e^++$H$^-$ channel in the CC model.

Before finishing the introduction, we note another relevant developement
in this subject. For the last few years we have been engaged in
addressing the Ps scattering problems by using a model exchange
correlation potential in the {\it ab initio} CC formalism. 
 Apart from the recent theoretical
development on Ps scattering by the Belfast and Calcutta groups employing
complete {\it ab initio} formalism, 
studies has contributions from two more
groups other than the Belfast group which has considered the
Ps-pseudostate technique \cite{wl0,wl1,wl2}, mentioned above. The
Calcutta group has addressed the Ps-H problem in various ways
\cite{hrg,sg1,sg2}, but they have not reported the binding or resonance
parameters. So, these aspects and hence the overall convergence of their
low energy scattering results could not be assesed comparing with the
accurate variational predictions. 

In the last few years, we 
are addressing Ps scattering problems by using a model regularized
exchange-correlation potential (non-local) \cite{ba1} in the exact
coupled Lippmann-Schwinger equation
\cite{cpl,nimb,pb1,bam2,ba1,ab1,abs,bam,ba0,bd}. In these calculations,
convergence and agreement with accurate variational predictions are
achieved by employing a minimum number of effective scattering channels
and tuning the exchange correlation potentials by means of a parameter.
The model agreed well with the measured data \cite{lr,gd,hy,rol} in Ps-He
\cite{}. When applied to Ps-H, it reproduces the precise binding and
resonances energies. Thereafter, it has laso been applied to other
taegets (H$_2$, Ne, Ar, Li) and agreement with measured data is obtained.
However, the crucial aspect of the model remains the tuning of the
parameter. It appears that the exchange correlation potentials, when
tuned, compensate for the correlation and continuum effects (see details
in the results and discussions). Thus, we expect that the direct
inclusion of correlation and continuum effects in the model CC framework,
through the H$^-$ formation channel, would diminish the role of the
parameter. Also, the charge transfer channel, in general, is expected to
improve the results obtained in the ab-initio framework due to possible
enhancement of correlation and continuum effects.

To have a preliminary assessment of these assertions, we re-investigate
the Ps-H scattering including the process Ps+H$\rightarrow e^+$+H$^-$. In
the present work, for a first estimation, we use a simple wave function
of Chandrasekhar for the H$^-$ ion \cite{hm}. Any elaborate wave function
containing the ${\bf r}_{12}$ term would certainly enhance the
electron-electron correlation further. We estimate the effect of this
charge-transfer recombination channel in the {\it ab initio} CC framework
by taking exchange exactly and also by using the regularized model
exchange \cite{ba1} in the CC equations.

{\bf  Theory:} Antisymmetrizing for the constituent electrons, 
we expand the total wave function of the Ps-H system in terms of Ps
$(\chi_\nu)$, H $(\phi_\mu)$, and H$^-$ ($\psi_\rho$) states as
\begin{eqnarray}
\Psi^{\cal S}({\bf r_1,r_2,x})&=& \sum_\nu \sum_\mu\biggr[
F_{\nu\mu}({\bf s_{2}})\chi_\nu ({\bf t_{2}})\phi_\mu({\bf
r_1}) +
(-1)^{\cal S}F_{\nu\mu}({\bf s_{1}})\chi_\nu ({\bf
t_{1}})\phi_\mu({\bf r_2})\biggr] \nonumber \\ &+&
\delta_{{\cal S}0} G_\rho({\bf x})\psi_\rho ({\bf
r_1,r_2}) ,
\end{eqnarray}
where ${\bf r_1, r_2}$ denote the electron
coordinates, and ${\bf x}$ is the
positron coordinate of Ps; ${\bf s_i= (r_i+x)}/2$, ${\bf t_i=
(r_i-x)}$; $i=1,2$. ${\cal S}$ is the total electron spin of the system
in a
particular channel which can have values $0$ and $1$,
corresponding to singlet and triplet scattering; $\delta_{{\cal S}0}$, is
the Kronecker delta.
$F_{\nu\mu}$ is the continuum orbital
of Ps and $G_\rho$ is that of the rearranged positron with respect to the
center of mass fixed in the target nucleus.
The total Hamiltonian of the system may be taken as
\begin{eqnarray}
H=-\frac{1}{4}\nabla_{Ps}^2+H_{Ps}^0+H_H^0+V{(1)}
\end{eqnarray}
or
\begin{eqnarray}
H=-\frac{1}{2}\nabla_{p}^2+H_{H^-}^0+V{(2)}
\end{eqnarray}
where $H_{Ps}^0$, $H_H^0$, and $H_{H^-}^0$ are the unperturbed Hamiltonian
for the Ps, H, and H$^-$, respectively; 
$-\frac{1}{4}\nabla_{Ps}^2$ and $-\frac{1}{2}\nabla_{p}^2$ represent the
kinetic energies of the Ps and the positron, respectively. 
$V{(1)}$ is
the interaction potential in channel-1 (elastic) and
$V{(2)}$ is the same in the
charge-transfer rearrangement channel (channel-2: $e^+$H$^-$). 
For channel-1, we have direct scattering (denoted by superfix $d$) and
exchange scattering (denoted by superfix $e$) and for
the
transfer channel, there is no notion of exchange and we have only the
direct scattering. The coulomb
potentials
for these channels are given by:
\begin{eqnarray}
V^d{(1)} &=& \frac{1}{x}-\frac{1}{|{\bf x-r_2}|}
-\frac{1}{r_1}+\frac{1}{|{\bf r_1-r_2}|}  \\
V^e{(1)} &=& \frac{1}{x}-\frac{1}{|{\bf x-r_1}|}
-\frac{1}{r_2}+\frac{1}{|{\bf r_1-r_2}|}  \\
V{(2)} &=& \frac{1}{x}-\frac{1}{|{\bf x-r_2}|}
-\frac{1}{|{\bf x-r_1}|}
\end{eqnarray}
We use exact wave function for Ps and H, and for H$^-$, we use the wave
function of Chandrasekhar \cite{hm}. 
Understanding the difficulty in dealing with a
coulomb wave for
the outgoing positron, we treat it with a plane wave in the input Born 
matrix element for channel-2, as the
distortion to this plane wave will partly be done through the use 
of dynamical Lippmann-Schwinger equation (given in the next
paragraph) and coupling over 
intermediate states.

Projecting over final states of Ps ($\chi_{\nu'}$), H ($\phi_{\mu'}$),
and H$^-$($\psi_\rho')$, the LS
equation for particular electronic spin state `${\cal S}$',
(which can have values $0$ and $1$, in general but for
$e^+$H$^-$ channel, it is ought to be $0$ only as H$^-$ can exist only
in singlet state), can be recast
as \cite{mgs}: 
\begin{eqnarray} \label{2} f^{\cal S}_{\nu'\mu',\nu\mu} ( {\bf
k_f,k_i})&=&
{\cal B}^{\cal S} _{\nu'\mu',\nu\mu }({\bf k_f,k_i}) \nonumber \\
&-&\frac{1}{2\pi^2} \sum_{\nu{''}}\sum_{\mu''} \int {d\bf k_1{''}}\frac
{{\cal B}^S _ {\nu'\mu',\nu''\mu{''}} ({\bf k_f,k_1{''}}) f^{\cal S}
_{\nu{''}\mu'',\nu\mu} ({\bf k_1{''},k_i})}
{k_{\nu{''}\mu{''}}^2-k_1{''}^2+i0} \nonumber \\
&-& \delta_{{\cal S}0}
\frac{1}{2\pi^2} \sum_{\rho{''}} \int {d\bf k_2{''}}\frac
{{\cal B}^{{\cal S}=0} _ {\nu'\mu',\rho{''}} ({\bf k_f,k_2{''}}) 
f^{{\cal S}=0}_{\rho{''},\nu\mu} ({\bf k_2{''},k_i})} 
{k_{\rho{''}}^2-k_2{''}^2+i0} 
\end{eqnarray} 
\begin{eqnarray} \label{2a} f^{{\cal S}=0}_{\rho',\nu\mu} ( {\bf
k_f,k_i})&=&
{\cal B}^{{\cal S}=0} _{\nu'\mu',\nu\mu }({\bf k_f,k_i}) \nonumber \\
&-&\frac{1}{2\pi^2} \sum_{\rho{''}}\int {d\bf k_2{''}}\frac
{{\cal B}^{{\cal S}=0} _ {\rho',\rho{''}} ({\bf k_f,k_2{''}}) 
f^{{\cal S}=0}_{\rho'',\nu\mu} ({\bf k_2{''},k_i})}
{k_{\rho {''}}^2-k_2{''}^2+i0} \nonumber \\
&-&\frac{1}{2\pi^2} \sum_{\nu{''}}\sum_{\mu''} \int {d\bf k_1{''}}\frac
{{\cal B}^{{\cal S}=0} _ {\rho',\nu{''}\mu''} ({\bf k_f,k_2{''}}) 
f^{{\cal S}=0}
_{\nu{''}\mu'',\nu\mu} ({\bf k_2{''},k_i})} {k_{\rho{''}}^2-k_2{''}^2+
i0} \end{eqnarray} 
where, $k_{\nu{''}\mu{''}}^2=\frac{2m_2}{\hbar
^2}\{E-\epsilon_{\nu''}-\Upsilon_{\mu''}\}$; and
 $k_{\rho {''}}^2=\frac{2m_1}{\hbar
^2}\{E-\upsilon_{\rho''}\}$; ${\bf k}_{\nu''\mu''}$ and ${\bf
k}_{\rho''}$ 
are the on-shell momenta; $m_1$ and $m_2$ are the  masses of
$e^+$ and Ps, respectively. $E$ represents the total energy of the
system; $\epsilon_{\nu''}$,
$\Upsilon_{\mu''}$, and $\upsilon_\rho{''}$ represent the binding
energies of Ps, H,
and H$^-$ ion, respectively. It is interesting to note that, for the
charge-transfer channel, the
summation over $\rho''$ is restricted by itself as without a magnetic
field, H$^-$ can exist only in its ground singlet state
(${\cal S}=0$) \cite{rnh}. We
study the effect of this channel over the elastic scattering channel
(channel-1) and
thus we restrict the summation over $\nu''$ and $\mu''$ to the ground
states of Ps and H, respectively. 
The input potentials to the coupled equations, in general, are given by
\begin{eqnarray}\label{st}
{\cal B}^{0,1}_{\nu'\mu', \nu\mu}({\bf k_f,k_i}) &=& 
B^{d}_{\nu'\mu', \nu\mu}({\bf k_f,k_i})+
(-1)^{0,1} B^{e}_{\nu'\mu', \nu\mu}({\bf k_f,k_i}) \nonumber \\
{\cal B}^{0}_{\rho, \nu\mu}({\bf k_f,k_i}) &=& 
B^{d}_{\rho, \nu\mu}({\bf k_f,k_i})
\end{eqnarray} 
Where, $B^d$ and $B^e$ are the direct Born and exchange Born-Oppenheimer 
(BO) amplitudes. The exact form for $B^d_{\nu'\mu', \nu\mu}$, is
available in
the literature
\cite{wl0,ba1} but here we follow the sign convention for the Ps-wave
function
as used in \cite{nimb}. For $B^e$, we
consider both the exact form (Appendix I) as well as the regularized
form \cite{ba1} with appropriate sign convention \cite{nimb}. We use
$\hbar=c=m_2=1$, where $m_2$ is the mass of electron or positron and
$m_1=2$, the mass of Ps.

The elastic cross section and H$^-$ formation cross section is given by:
\begin{eqnarray}
\sigma_{el} &=& \frac{1}{4}|f_{1s 1s,1s 1s}^{0}|^2 + 
\frac{3}{4}|f_{1s 1s,1s 1s}^1|^2 \\
\sigma_{H^-} &=& \frac{|k_f|}{|k_i|}\frac{1}{4}|f_{1s^2,1s 1s}^{0}|^2  
\end{eqnarray}

Recently, a concern has been raised by Adhikari and Mandal (AM) \cite{am}
regarding the validity of the calculation of the BO exchange term of
Sinha et al \cite{sg1} and provided a different results for this BO
exchange term. We were also concern with AM regarding the feature of a
minimum in the elastic cross section with BO exchange near 35 eV.
However, present results for $B^e_{1s1s,1s1s}$ exchange agree exactly
with those of Sinha et al \cite{sg1} and when this BO amplitude is used
in the coupled equations, along with the required off-shell matrix
element, it reproduces the static-exchange phase shifts of Campbell et al
\cite{wl1}, Sinha et al \cite{sg1}, and Hara and Fraser \cite{hf} quite
precisely (discussed in the results and discussion).  So, it implies that
the BO exchange amplitudes in these calculations are same and the minimum
in the cross section is a feature of the BO exchange term and not due to
any error in its evaluation. AM evaluates the BO exchange term relying
mostly on numerical calculations where the integrand contains spherical
Bessel functions and they have reported to use 400 Gauss-quadrature
points in the evaluation of the integrals. Whereas, using Fourier
transform and Chasire integrals, we reduce the nine-dimensional integrals
(Appendix I) to simple two-dimensional integrals which is found to
converge with mere eight Gauss-quadrature points for each integration
variable and hence any numerical problem is ruled out in this methodology
compared to that provided by AM \cite{am}.

In Appendix I, we provide the analytical form of the {\it ab initio} BO
exchange matrix element $B^e_{1s 1s, 1s 1s}$ and the charge-transfer
matrix element $B^d_{1s^2, 1s 1s}$ is detailed in Appendix II. In a
separate study, we employ the regularized model exchange for $B^e_{1s 1s,
1s 1s}$, which is taken from ref.\cite{nimb} (it differs only in sign
convention from ref.\cite{ba1}).

{\bf Numerical Procedures :} In the present investigation, we consider
static-exchange (SE) and SE plus charge-transfer reactions of Ps-H to
$e^+$-H$^-$. The three-dimensional LS equations, for a particular
electronic-spin state ($S$) are decomposed to coupled one-dimensional
partial wave equations, which are then solved by the method of matrix
inversion. We find that at low energies, 44-48 Gauss
quadrature points (24 points for $k_{1}''=0-2k_{\nu''\mu''}$ interval and
20-24 points for
$k''=2k_{\nu''\mu''}-\infty$ interval and similarly for $k_2''$ are
needed for the discretization of the Kernel of the
LS equation, to achieve numerical convergence up to 4-decimal places.

{\bf Results and Discussions :} We begin the discussions of our results
with the singlet channel cross sections.  The effect of the
Ps+H$\rightarrow e^+$+H$^-$ channel would be revealed in this singlet
channel, as H$^-$ belongs to a spin singlet state only.  Also, the
binding and resonance of PsH occur in this channel.  In figure-1, we plot
the S-wave singlet cross section from 0-10 eV to exhibit the influence of
this channel (channel-2) over the static-exchange prediction
(using both exact and
model exchange). In both the cases, we find that this channel reduces the
singlet scattering cross section significantly, at low energies. This
implies that this channel effectively makes the attractive potential more
strong and we can expect a smaller scattering length and a greater
binding energy. Using an effective-range expansion of the form
$k$cot$\delta=-1/a+r_0k^2/2+Dk^4$, and finding solution of the equation
$k$cot$\delta-ik=0$ for the bound state, where $k$ is the momentum of Ps,
$\delta$ is the S-wave singlet phase shift, $r_0$ is the corresponding
effective range, and $D$ is the coefficient for the $k^4$ term, we obtain
the binding energy and scattering length which are tabulated in table 1.

{\small {\bf Table 1}: Scattering length (S.L.) and binding energies
(B.E.) in the singlet channel of Ps-H, from static-exchange (SE),
static-exchange plus charge-transfer rearrangement (2CH) employing exact
exchange. Model exchange results without any parameterization ($C=1$) are
represented by (ME). a) present
predictions; b) prediction of Campbell et al \cite{wl1}}

\begin{center}
\begin{tabular}{l l l l l l }
\hline
   & SE & 2CH & SE(ME) & 2CH(ME) & 22-state  \\
\hline
  S.L. & a) 7.273 & 6.90  & 8.427 & 5.544 &   \\
       & b) 7.25  &       &       &       & 5.20   \\
\hline
  B.E. & a)-0.253 &-0.291 &-0.139 &-0.394 &     \\
       & b)-0.263 &       &       &       &-0.634   \\
\hline
\end{tabular}
\end{center}
\vskip 10pt

As expected, we find the scattering length (S.L.) is reduced and the
binding energy (B.E.) is increased with 
the influence of the H$^-$ formation channel. The effect is much
pronounced in the case of
model exchange calculation, thus signifying a lesser role for the
parameter in the model potential. The scattering length prediction in
the 2CH(ME) is quite close to the 22-state prediction. Keeping in mind
the simple form of H$^-$ wave function used in this calculation, and the
proven role of the direct ${\bf r_{12}}$ term in the variational
prediction \cite{ho1}, we get the impression that the impact of this
channel in the {\it ab initio} model could be improved significantly by
using a more shopisticated wave function for H$^-$ which contains the
${\bf r_{12}}$ term. The small difference in the scattering length
(S.L.) and binding energy (B.E.) of SE calculations of set (a) and set
(b) is due to the fact that in (b) we have used an effective range
expansion to extrapolate the phase shifts to the negative energy region
while in set (a) those parameters are directly solved. The
positive energy scattering phase shifts for (a) and (b) agree exactly.
We tabulate the phase shifts for different models in table 2 for
future reference.

{\small {\bf Table 2}: Variation of singlet and triplet scattering phase
shifts for different models with energy ($E=6.8k^2$ eV): SE-static
exchange; 2CH-static exchange plus $e^+H^-$ channel; (ME)-represents the
calculations with model exchange but without any parameterization ($C=1$).

\begin{center}
 \begin{tabular}{ l l l l l|l l }
\hline
 & &  singlet &  & &  triplet  & \\
\hline
   $k$  &  SE & 2CH & SE(ME) & 2CH(ME) & SE & SE(ME)  \\
\hline
  0.1  & 2.455 & 2.485 & 2.387 & 2.610 & -0.247 & -0.145  \\
  0.2  & 1.927 & 1.966 & 1.910 & 2.176 & -0.489 & -0.283  \\
  0.3  & 1.539 & 1.577 & 1.604 & 1.848 & -0.721 & -0.410  \\ 
  0.4  & 1.239 & 1.275 & 1.381 & 1.596 & -0.940 & -0.521  \\ 
  0.5  & 0.997 & 1.032 & 1.207 & 1.396 & -1.143 & -0.613  \\ 
  0.6  & 0.797 & 0.831 & 1.065 & 1.232 & -1.330 & -0.683  \\ 
  0.7  & 0.631 & 0.664 & 0.946 & 1.094 & -1.499 & -0.731  \\ 
  0.8  & 0.491 & 0.523 & 0.845 & 0.977 & -1.653 & -0.757  \\ 
  0.9  & 0.373 & 0.406 & 0.758 & 0.876 & -1.790 & -0.763  \\ 
  1.0  & 0.274 & 0.307 & 0.683 & 0.786 & -1.913 & -0.781  \\ 

\hline
\end{tabular}
\end{center}
\vskip 10pt

First we discuss about SE phase shifts where recently a dispute has been
raised \cite{am}. Table 2 shows that the present phase shifts for the SE
calculation for the singlet and triplet scattering agree exactly with the
existing SE predictions of Hara and Fraser \cite{hf} and Sinha et al
\cite{sg1}.  Campbell et al \cite{wl1} provided SE phase shifts for other
$k-$values (Ps energy, $E=6.8k^2$). At $k^2=0.1639, 0.2478,$ and $0.5588$
({\small a.u}) the singlet phase shifts of Campbell et al are given by
(table 1 of ref. \cite{wl1}) 1.23, 1.00, 0.56, respectively and our
predicted values are 1.228, 1.002, and 0.5614, respectively (all cross
sections are in units of $\pi a_0^2$). So all these SE predictions, which
use BO exchange amplitude as the input, agree among themselves. This
suggests that the minimum in the BO cross section is a true feature of
the model and the results of Adhikari and Mondal (AM) \cite{am} cannot
agree with Campbell et al \cite{wl1}. Since, at $k^2=0.5588$, the BO
cross sections of AM \cite{am} is about $23.5 \pi a_0^2$, which is
approximately 22\% higher than the BO cross sections ($19.3 \pi a_0^2$)
of Sinha et al and present calculation. So, it is certain that any static
exchange prediction made using the BO exchange amplitude of AM as input
to the CC equations, will disagree with all existing SE results including
those of Campbell et al \cite{wl1}.  While trying to get an answer to the
minimum in the elastic BO cross section, we find that for forward
scattering the exchange potential changes its sign at about 26-27 eV,
which is causing the minimum in the cross section. We now come to the
main business of the asssement of the 
influence of the H$^-$ formation channel over the SE predictions.

In table 2, we compare the phase shifts of SE and 2CH calculations. For
2CH we do not tabulate the triplet phase shifts since they do not change
by the influence of channel-2 (as H$^-$ exists only in singlet state).
From table 2, we see that singlet phase shifts increase significantly
with the inclusion of the H$^-$ formation channel. This signifies an
increase in the attractive potential, which was expected and needed in
the CC theory to improve the convergence and hence improve the PsH
binding energy. It is interesting to note that the effect of this channel
continues significantly around $k=1.0$.

In figure 2, we plot the elastic cross sections for SE, SE(ME), 2CH, and
2CH(ME) models. In this figure we also plot the recent variational
prediction \cite{am2} on the zero energy cross section. Elastic cross
sections for both the {\it ab initio} and model calculations are reduced
by the influence of the charge transfer channel. At low energies, the
2CH(ME) model gives much lower cross section than the 2CH model and the
former is quite close to the recent variational prediction \cite{am2}. We
have earlier find that this model can lead to more converged results if
the model exchange potential is tuned by means of a parameter $C$
\cite{cpl,ba1}. While trying to analyze the physics behind such agreement
with measurements and accurate variational predictions, we find that
effectively, while making approximate mapping of a basis set
($\psi_\mu({\bf r}_2)$)  belonging to a different Fock-space to its
original Fock-space ($\psi_\mu({\bf r}_1)$), basically we are indulging
to approximations like: \begin{eqnarray} \int \psi_\mu({\bf
r}_2)\frac{1}{|{\bf r}_1-{\bf r}_2|} d{\bf r}_2 &=& \frac{1}{Ck_\mu^2}
\psi_\mu({\bf r}_1) \\ \int \phi_\nu({\bf x-r_2}\frac{1}{|{\bf r}_1-{\bf
r}_2|} d{\bf r}_2 &=& \frac{1}{Ck_\nu^2} \phi_\nu({\bf x-r_1})
\end{eqnarray} where $k_\mu^2$ and $k_\nu^2$ are the average values of
the square of the momenta of the respective electrons when they were
bound in the hydrogen ($\psi_\mu({\bf r}_2)$) or in the positronium atom
($\phi_\nu({\bf x-r_1})$) and $C$ is a parameter which is fixed to unity
but could be varied to tune results. Clearly, the above relations are
introducing some amount of $e_1-e_2$ correlation and the continuum
effects of the target (eqn.()) and the projectile (eqn.()), respectively.
Also, these effects would be modified when the value of $Ck_\mu^2$ or
$Ck_\nu^2$ are changed. Interestingly, the model provides best results
\cite{cpl,ab1,nimb} when $Ck_\mu^2$ and $Ck_\nu^2$ become close to the
ionization energies for the respective electrons. Although the mixing of
such model exchange with {\it ab initio} charge-transfer process yielded
very good response, we are doubtful about the overcompleteness of the
Hilbert space if such mixing is continued with a larger basis for Ps.  
However, from the results of the 2CH model and from our experience in
such rearrangement for positron-atom scattering \cite{}, we understand
that for the full {\it ab initio} model (2CH) the possibility of
overcompleteness is remote unless a very large basis is used. The
information we gain here in terms of the convergence of the 2CH
prediction appears consistent and encouraging due to the fact that
virtual excitations of Ps are absent here. The charge-transfer reaction
becomes exothermic for $n\ge 5$ and higher discrete excited states and
continuum of Ps and hence the effect of such a channel is likely to be
significant with higher Ps eigen states and pseudostates. Consideration
of this channel along with the Ps-excited states would also reveal its
effect on resonances, which manifest in the theory with the inclusion of
Ps excited states.

Another interesting feature of such study is that it provides the H$^-$
formation cross section which is of importance in Astrophysics. Here we
provide the H$^-$ formation cross sections from Ps(1s)-H(1s) scattering
in figure-2. Solid curve is obtained with the regularized exchange
\cite{ba1,nimb} while the dotted curve is obtained with {\it ab initio}
exchange. It is observed that from 60 eV onwards the two sets of results
are same signifying the fact that the regularized model exchange
asymptotically coalesce with the exact one. From the two curves we also
see that, in the low and intermediate energies, the cross sections
calculated with the regularized exchange CC model (2CH(ME)) are higher
than those calculated with {\it ab initio} exchange (2CH). This was quite
expected as the regularized exchange potential, as discussued above, was
found to simulate the continuum effect substantially and yield converged
results for the PsH binding and resonances \cite{cpl,ab1}. The
contribution from virtual Ps excitations, no doubt will increase the
H$^-$ formation cross section and this trend has been revealed in the
2CH(ME) calculation.

{\bf Conclusion :} In summary, we make a first time study of the effect
of charge transfer recombination for Ps scattering from hydrogen.
Considering the rearrangement of PsH to $e^+$H$^-$ we get the information
that such rearrangement can introduce the continuum effect more
effectively than considering only exchange. The attractive potential
introduced by this channel improves the scattering length and binding
energy predictions significantly. The influence of the channel is also
found to help converging the low energy cross sections (see figure 1).
Similar effect of charge-transfer recombination was found in the cases of
positron-atom scattering \cite{msa,msb} and in the present case of
Ps-atom scattering it could be more interesting as the threshold for the
transfer process Ps(1s)+H(1s)$\rightarrow e^++$H$^-$ decreases with
higher Ps states and the reaction becomes exothermic for $n\ge 5$
discrete excited states and continuum of Ps. So, it is expected that the
channel would contribute substantially through virtual Ps excitations and
lead to further convergence for the low energy Ps scattering parameters.
As a by product we get H$^-$ formation cross sections which has
astrophysical importance.

The work is  supported by the Funda\c c\~ao de Amparo \`a Pesquisa
do Estado de S\~ao Paulo of Brazil via project number 99/09294-8.

\newpage

\begin{center}
{\large Appendix I}
\end{center}
 
The Born-Oppenheimer exchange matrix element for the elastic transition
is given by:

\begin{eqnarray} \label{9} 
B^e_{1s 1s, 1s 1s} ({\bf k_f, k_i}) =  -\frac{m_{1}}{2\pi} 
\int\int\int & & e^{-i\frac{1}{2}{\bf k}_f.({\bf r}_2+{\bf x})}
\chi_{1s}({\bf r}_2-{\bf x})\phi_{1s}({\bf r}_1) V_{int}^{(1)} 
\nonumber \\
&\times & e^{i\frac{1}{2}{\bf k}_i.({\bf r}_1+{\bf x})}
\chi_{1s}({\bf r}_1-{\bf x})\phi_{1s}({\bf r}_2) d{\bf x}d{\bf r}_1d{\bf
r}_2.
\end{eqnarray} 

where ${\bf
k}_i$, ${\bf k}_f$ are the initial and final momentum of the Ps atom with
respect to the center of mass. We use $\hbar=c=m_2=1$, $m_2$ is the mass
of
the electron or positron; the Ps mass $m_{1}=2$. Splitting with the
four terms of $V_{int}^{(1)}$ of
eqn(4), we write $B^e_{1s 1s, 1s 1s} = B_1 + B_2 + B_3 + B_4$, where
subscripts denote the terms on the {\it r.h.s.} of eqn(4), in
sequence. We
use Fourier transform, delta-function integration, and Chasire integrals
to reduce each nine
dimensional integrals to two dimensions. 
We follow same procedure for $B_1,
B_2, B_3, B_4$. We detail for $B_4$, which contains $1/|{\bf r_1-r_2}|$
term of
$V_{int}^{(1)}$. 
Using $\chi_{1s}(r)=\frac{1}{{\sqrt {8 \pi}}}e^{-\beta r}$
and $\phi_{1s}(r)=\frac{1}{{\sqrt \pi}}e^{-\alpha r}$, with $\beta=0.5$
and $\alpha=1.0$, 
we perform integration over $d{\bf x}$ first,
then over $d{\bf r_1}$, and finally over $d{\bf r_2}$ to get,
\begin{eqnarray}
B_4=4\beta^2\alpha\int_0^1 du
u(1-u)\int_0^1 dv v 
\biggr(\frac{1}{\mu_1}\frac{\partial}{\partial\mu_1}\biggr)^2\frac{1}{\mu_1}
\biggr(\frac{1}{\mu_2}\frac{\partial}{\partial\mu_2}\biggr)\frac{1}{\mu_2}
\frac{\mu_{2\alpha}}{(|{\bf \Lambda}|^2+\mu_{2\alpha}^2)^2}
\end{eqnarray}
where
\begin{eqnarray}
\mu_1^2 &=& \beta^2+u(1-u)|0.5({\bf k_i-k_f})|^2 \\
\mu_2^2 &=& v\alpha^2+(1-v)\mu_1^2+v(1-v)|{\bf Q}|^2 \\
{\bf Q} &=& (1-0.5u){\bf k_i} - 0.5(1-u){\bf k_f} \\
{\bf \Lambda} &=& (0.5u+v-0.5uv){\bf k_i} - 0.5(1+u+v-uv){\bf k_f} \\
\mu_{2\alpha} &=& \mu_2+\alpha
\end{eqnarray}
For the rest of $B_1, B_2$ and $B_3$, we perform integration over
$d{\bf r_1}$ first, then over $d{\bf x}$, and finally over
$d{\bf r_2}$ and arrive at similar final forms as
above, with changed definition for $\mu_2$, ${\bf Q}$, ${\bf \Lambda}$
etc. 
Integration
over each of $du$ and $dv$ is converged with  mere eight
Gauss-quadrature
points.

\newpage

\begin{center}
{\large Appendix II}
\end{center}
 
The charge transfer matrix element for the
Ps+H$\rightarrow e^+$+H$^-$ channel is given by:

\begin{eqnarray} \label{10} 
B^d_{1s^2, 1s 1s} ({\bf k_f, k_i}) =  -\frac{m_{1}}{2\pi} 
\int\int\int  e^{-i{\bf k}_f.{\bf x}}
\psi_{H^-}({\bf r}_1,{\bf r}_2) V_{int}^{(2)} 
 e^{i\frac{1}{2}{\bf k}_i.({\bf r}_1+{\bf x})}
\chi_{1s}({\bf r}_1-{\bf x})\phi_{1s}({\bf r}_2) d{\bf x}d{\bf r}_1d{\bf
r}_2.
\end{eqnarray} 
where ${\bf k}_f$, now represents the momentum of the ejected 
positron with respect to the center of mass rest on the
proton; $m_1=1$; 
$\psi_{H^-}({\bf r_1,r_2})=\frac{1}{N}\biggr
(e^{-ar_1}e^{-br_2} + e^{-ar_2}e^{-br_1} \biggr)$, the H$^-$ wave
function
with binding energy
$E=-0.51330$ and
$a=1.03925$, $b=0.28309$ and the normalization constant has been
worked out to be $N=31.80348105$.
As in Appendix I, we split the integral as
$B^d_{1s^2, 1s 1s} = B_1 + B_2 + B_3 $, 
 with the three terms of $V_{int}^{(2)}$ of eqn(5),
where
subscripts denote the terms on the {\it r.h.s.} of eqn(5), in
sequence. Using the symmetry of the H$^-$ wave function we write
$B_j=B_j(a,b)+B_j(b,a); j=1,2,3.$
We
use Fourier transform, delta-function integration and Chasire integrals
and evaluate $B_2$ exactly.
\begin{eqnarray}
B_2(a,b)=-\frac{64{\sqrt 2}\pi^2 a}
{(\alpha+b)^3(|{\bf q}_{1/2}|^2+\beta^2)^2+(|{\bf q}|^2+a^2)^2}
\end{eqnarray}
where ${\bf q}_{1/2}=\frac{1}{2}{\bf k_i-k_f}$
and ${\bf q}={\bf k_i-k_f}$.
Integration over $d{\bf
r}_2$ in $B_1$ is simple and in $B_3(a,b)$, it
gives
\begin{eqnarray}
I=\int e^{-(\alpha+b)r_2}\frac{d{\bf r_2}}{|{\bf x-r_2}|}=
\frac{8\pi}{(\alpha+b)^3 x}-\frac{8\pi e^{-(\alpha+b)x}}{(\alpha+b)^3 x}
-\frac{4\pi e^{-(\alpha+b)x}}{(\alpha+b)^2}
\end{eqnarray}
With this form of $I$, we write
$B_3(a,b)=B_3^1(a,b)+B_3^2(a,b)+B_3^3(a,b)$. Now 
$B_3^1(a,b)$ cancels out with $B_1(a,b)$ and similarly for $B_3^1(b,a)$
and $B_1(b,a)$. We evaluate
$B_3^2(a,b)$ and obtain $B_3^3(a,b)$ from it with a derivative w.r.to
($\alpha+b$). The final form of $B_3^2(a,b)$ is
\begin{eqnarray}
B_3^2(a,b)=-\frac{32{\sqrt 2}\pi^2 \beta}{(\alpha+b)^3}
\int_0^1 du u 
\biggr(\frac{1}{\mu}\frac{\partial}{\partial\mu}\biggr)\frac{1}{\mu}
\frac{\mu+a}{\{|{\bf Q}|^2+(\mu+a)^2\}^2}
\end{eqnarray}
where
\begin{eqnarray}
\mu^2 &=& u\beta^2+(1-u)(\alpha+b)^2+u(1-u)|{\bf q}_{1/2}|^2 \\ 
{\bf Q} &=& (1-0.5u){\bf k_i} - (1-u){\bf k_f} 
\end{eqnarray}

\newpage


\newpage

{\bf Figure Captions:}

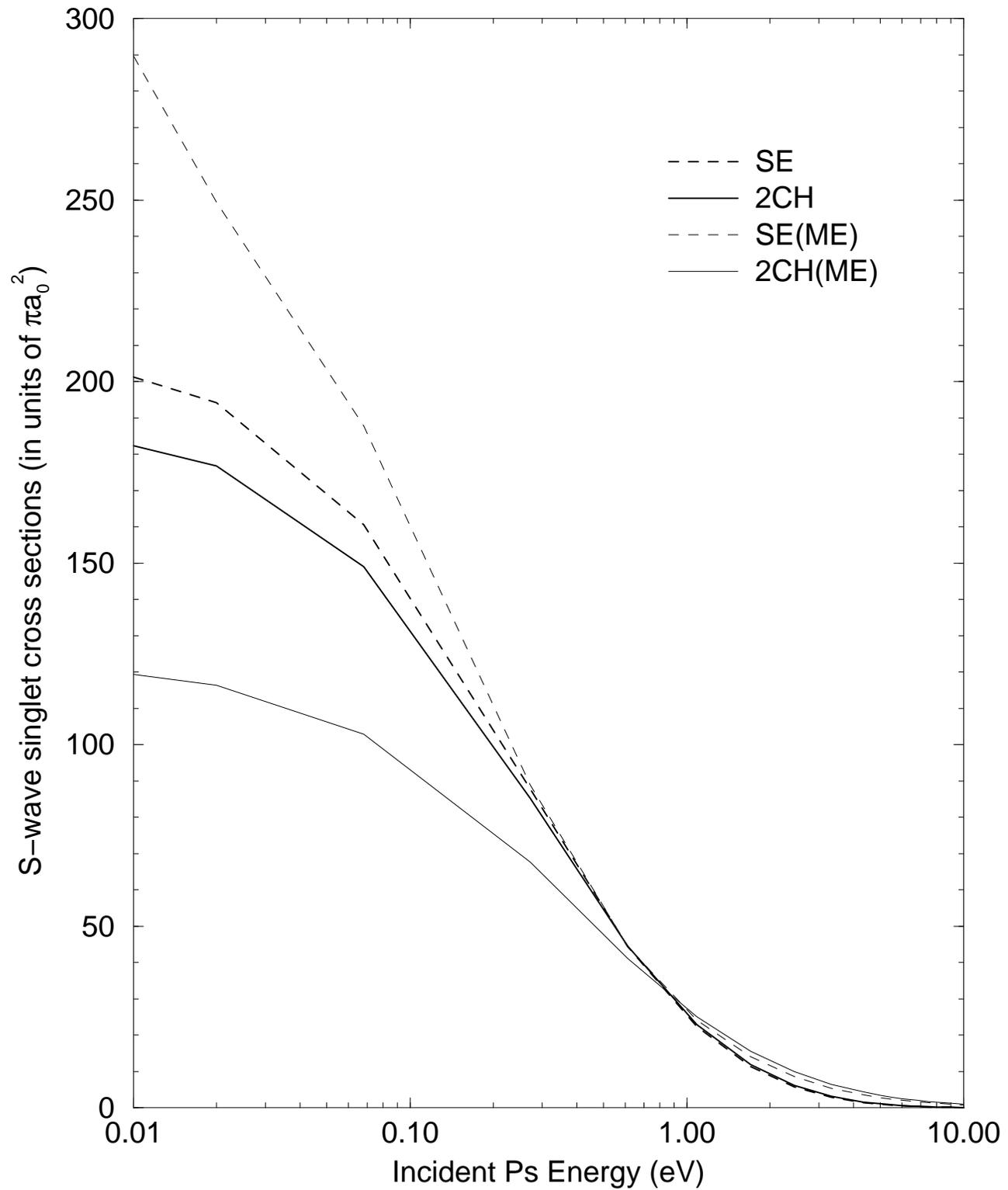
{\bf Fig. 1:} Variation of S-wave singlet cross sections (in units of
$\pi a_0^2$) for Ps-H scattering employing static-exchange (SE), SE plus
charge-transfer rearrangement to $e^+$H$^-$ (2CH). (ME) represents
calculations using model exchange \cite{ba1}:

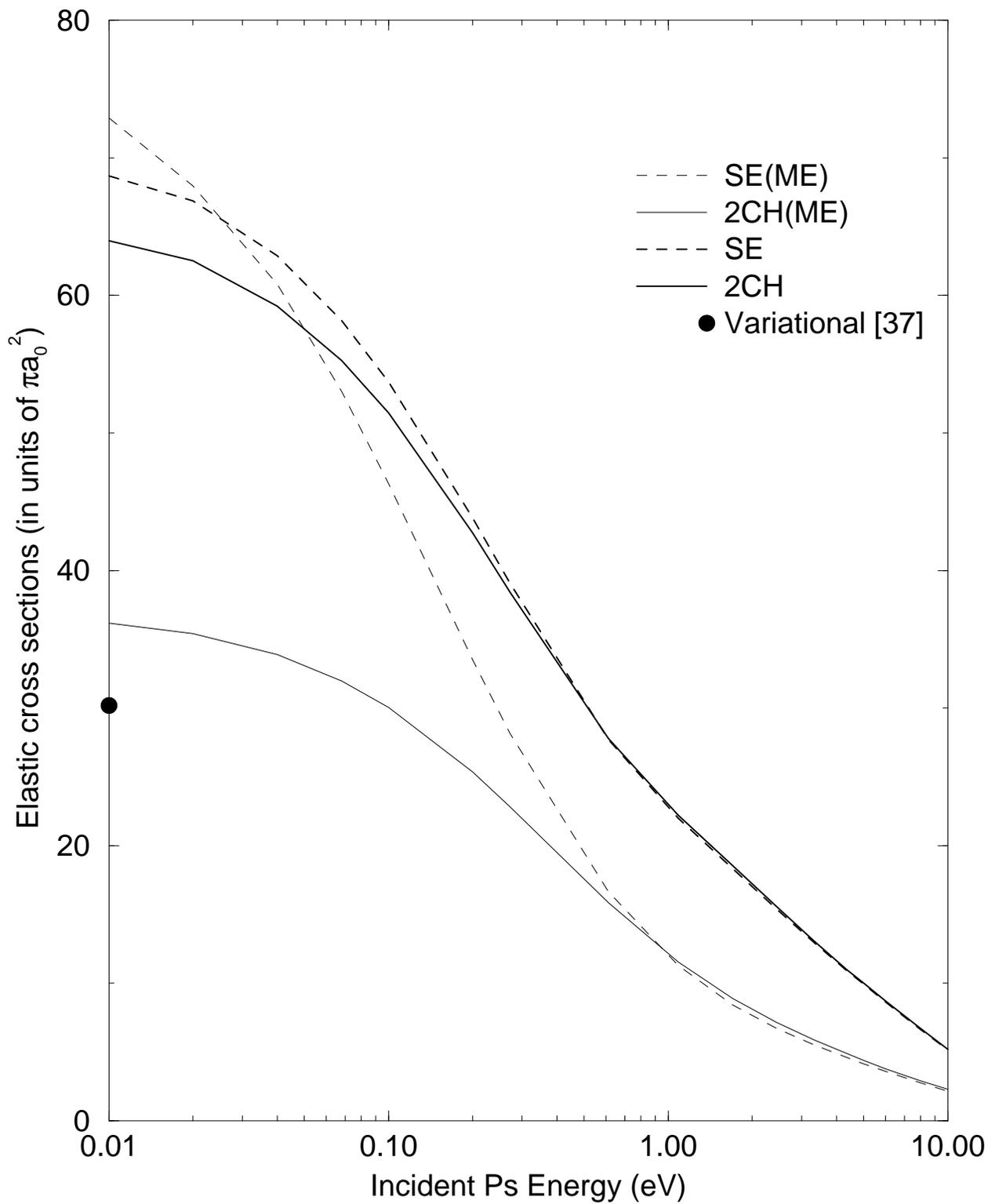
{\bf Fig. 2:} Variation of elastic cross sections (in units of
$\pi a_0^2$) for Ps-H scattering employing static-exchange (SE), SE plus
charge-transfer rearrangement to $e^+$H$^-$ (2CH). (ME) represents
calculations using model exchange \cite{ba1}:

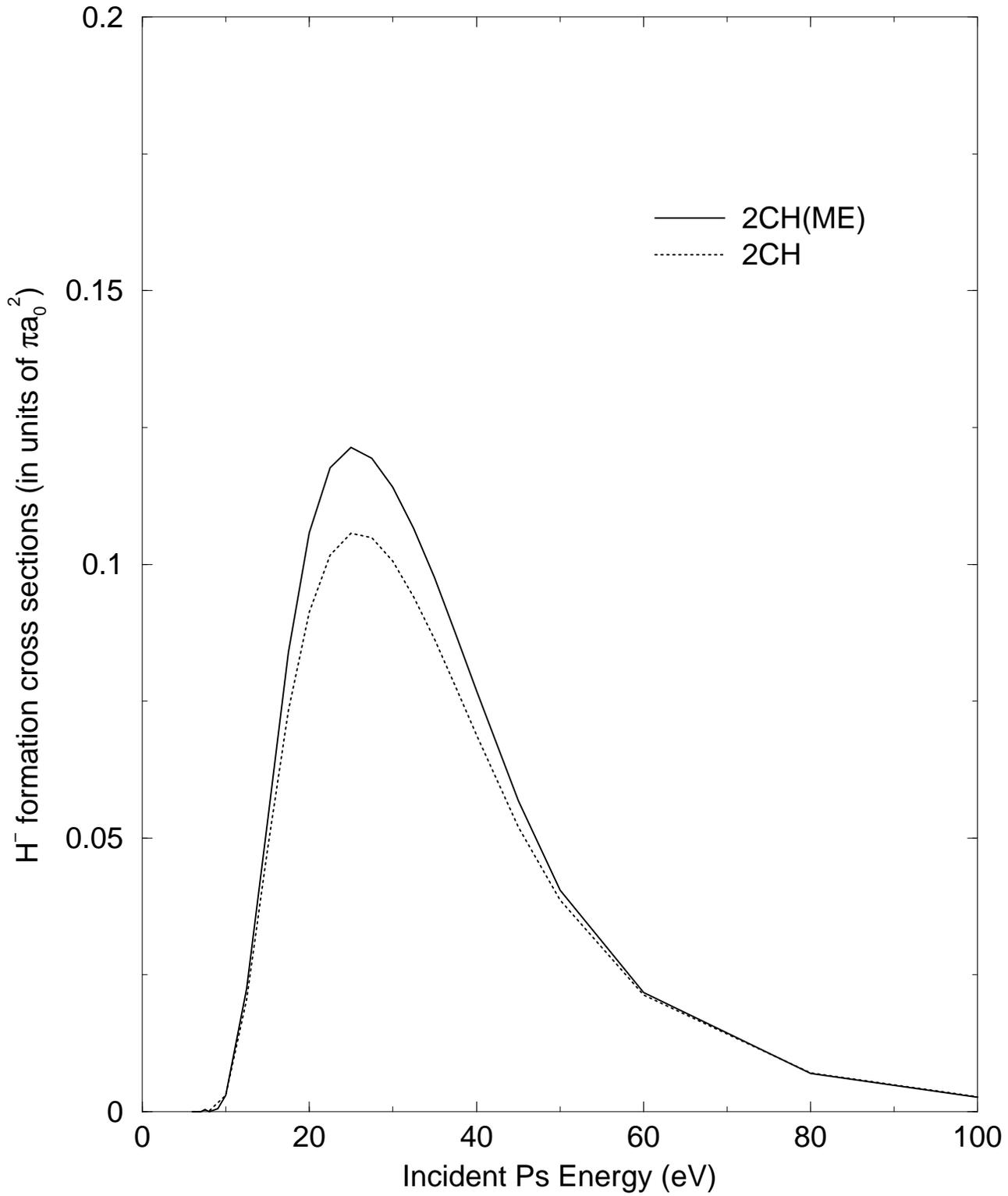
{\bf Fig. 3:} H$^-$ formation cross sections from Ps(1s)-H(1s) scattering
(in units of $\pi a_0^2$). 2CH corresponds to results fron the two-channel
calculation with {\it ab initio} exchange while 2CH(ME) is the same using
regularized model exchange.

\end{document}